\documentclass[a4paper,11pt]{article}
\usepackage{pos}
\usepackage[export]{adjustbox}

\title{Majorana neutrino dipole moments and masses at a muon collider}

\author[a]{Michele Frigerio}
\author*[b]{Natascia Vignaroli}

\affiliation[a]{Laboratoire Charles Coulomb (L2C), University of Montpellier, CNRS, Montpellier, France}
\affiliation[b]{Dipartimento di Matematica e Fisica E. De Giorgi, Universit\`a del Salento and INFN Sezione di Lecce, via per Arnesano 73100 Lecce (LE), Italy}

\emailAdd{natascia.vignaroli@unisalento.it}
\emailAdd{michele.frigerio@umontpellier.fr}

\abstract{A future multi-TeV muon collider would provide an important probe for Majorana neutrinos. A muon collider with a collision energy of $\sim$30 TeV would be sensitive to $\nu_e-\nu_\mu$ transition dipole moments of the order of $\sim 10^{-12}\mu_B$ and would be thus competitive with the latest astrophysical observation and laboratory experiments. Contrary to the latter, the muon collider would have the unique advantage of a direct and clean identification of lepton number and flavour violation. This would establish the Majorana nature of the neutrino and would provide crucial complementary information on the neutrino properties in the event of a (near) future observation at low-energy experiments. Additionally, a  muon collider would improve by orders of magnitude the direct bounds on the Majorana neutrino mass matrix entries $m_{e\mu}$ and $m_{\mu\mu}$. }

\FullConference{12th Neutrino Oscillation Workshop (NOW2024)\\
 2-8, September 2024\\
Otranto, Lecce, Italy\\}

\begin{document}
\maketitle

\section{Introduction}

We highlight in this proceeding the main results of the study in \cite{Frigerio:2024pvc}. Majorana neutrinos may have lepton number violating (LNV) and flavour non-diagonal transition dipole moments (TDMs) which, in compelling new physical scenarios, can reach values of the order of $\lambda_\nu \sim 10^{-12} \mu_B$, with $\mu_B\equiv e/(2m_e)$ the Bohr magneton, and thus become accessible experimentally. The current most stringent limits on neutrino TDMs are placed by cosmological and astrophysical observations, $\lambda_\nu \lesssim 10^{-12} \, \mu_B$  \cite{Capozzi:2020cbu}, subject, however, to large uncertainties, and by low-energy scattering experiments, with 90\% C.L. bounds of the order of $\lambda_\nu \lesssim 6 \times 10^{-12} \mu_B$ from XENONnT \cite{XENON:2022ltv} and LUX-ZEPLIN \cite{LZ:2022lsv}. The study in \cite{Frigerio:2024pvc} explores for the first time the sensitivity 
to TDMs of collider experiments, focusing in particular on the promising prospects of a future muon collider \cite{Accettura:2023ked}. The LNV signatures considered to probe neutrino TDMs at colliders are also sensitive to Majorana neutrino masses.

\section{Majorana neutrino dipole moments and masses at a muon collider}

Majorana neutrino TDMs \cite{Schechter:1981hw,Nieves:1981zt,Shrock:1982sc, Kayser:1982br} can be described by five-dimensional operators,  
\begin{equation}\label{eq:NMM-A}
\begin{array}{c}
{\cal L} \supset \dfrac{1}{4} \, \overline{(\nu_{L\alpha})^c} \sigma^{\mu\nu} \lambda_{\alpha\beta} \nu_{L\beta} A_{\mu \nu} + \text{h.c.}
\end{array}
\end{equation}
where $\nu_L$ are left-handed spinors, $\nu\equiv \nu_L + (\nu_L)^c $ are Majorana spinors,
$\sigma^{\mu\nu}\equiv (i/2)[\gamma^\mu,\gamma^\nu]$,
$A_{\mu\nu}$ is the photon field strength,
and $\alpha,\beta$ are flavour indexes. 
The TDM matrix  $\lambda$ has mass-dimension minus one, and it is antisymmetric in flavour space, $\lambda_{\alpha\beta}=-\lambda_{\beta\alpha}$. 

TDMs can be efficiently probed at a muon collider by analyzing the
 $\Delta L=2$ process in Fig.~\ref{fig:NMM-muCol}, which is induced directly by the TDM operator in Eq.~\eqref{eq:NMM-A}. The channel is characterized by a final state with two same-sign and different-flavour leptons, accompanied by two same-sign $W$'s (which is convenient to consider decaying hadronically). Cross sections (in Fig.~\ref{fig:NMM-muCol}) are orders of magnitude larger than those for analogous processes at current and future hadron colliders \cite{Frigerio:2024pvc}. The background is reducible and can be attenuated to an almost negligible level by exploiting the peculiar kinematics of the signal \cite{Frigerio:2024pvc}. The projected $2\sigma$ sensitivities on a $\nu_e-\nu_\mu$ TDM are reported on Table \ref{sensMU}.

The same LNV signatures considered to probe neutrino TDMs at colliders are also sensitive to Majorana neutrino masses,
\begin{equation}\label{MajM}
{\cal L}\supset -\dfrac 12 \overline{(\nu_{L\alpha})^c} \, m_{\alpha\beta} \, \nu_{L\beta}+\text{h.c.}\,,
\end{equation}
with $m$ a complex symmetric $3\times 3$ matrix in flavour space. Representative leading Feynman diagrams for the process $\mu^+\mu^- \to \ell_\alpha^\pm \ell_\beta^\pm W_h^\mp W_h^\mp$ induced by Majorana masses, and the corresponding cross sections at a muon collider are shown in Fig.~\ref{fig:NMM-muCol}. Table \ref{tab:sign-masses} reports the muon collider $2\sigma$ sensitivity limits on the Majorana mass matrix entries $|m_{e\mu}|$ and $|m_{\mu\mu}|$ obtained from the analysis in \cite{Frigerio:2024pvc}. Notice that current bounds at LHC, $|m_{e\mu,\mu\mu }|\lesssim$10 GeV, are weaker by far.

\begin{figure}
\centering
 \includegraphics[scale=0.6, valign=t]{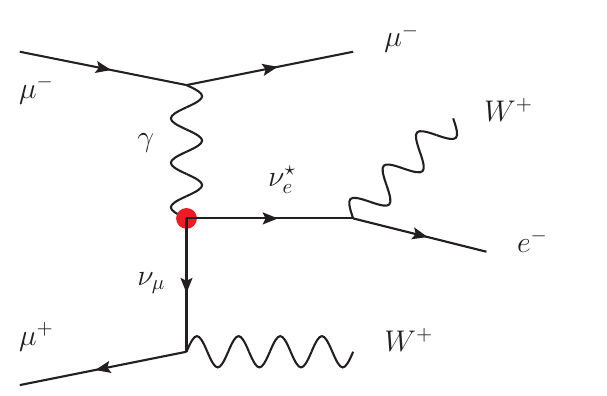} \includegraphics[scale=0.6, valign=t]{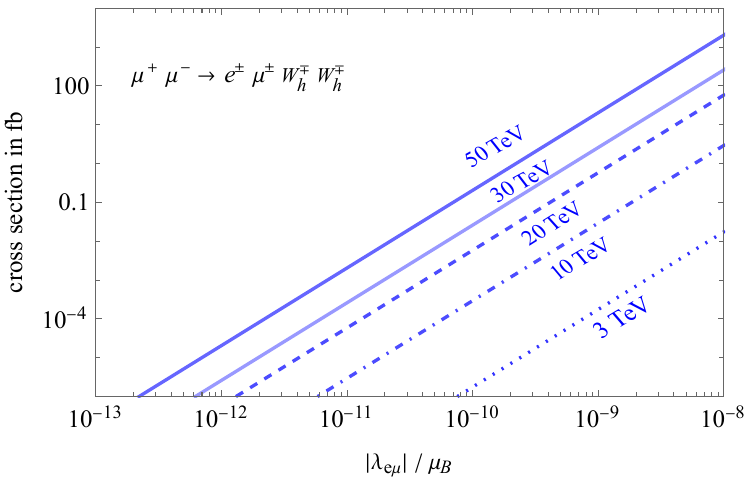}
 \caption{\em Leading Feynman diagram for the $\Delta L=2$ process induced by an electron-muon neutrino dipole moment and the corresponding cross sections at a muon collider.}\label{fig:NMM-muCol}
 \end{figure}

\begin{table}[]
\centering
{\small
\begin{tabular}{|c|ccccc|}
\hline 
 & & & & &  \\
 & & & & &  \\[-0.6cm]
\textsf{$\sqrt{s}$} & 3 TeV & 10 TeV & 20 TeV & 30 TeV & 50 TeV \\
  \hline \\[-2mm]   { $\frac{|\lambda_{e\mu}|}{ \mu_B} $}  & 5.4\;[4.8] $\cdot$ 10$^{-9}$ & 1.5\;[1.1]  $\cdot$ 10$^{-10}$ &  1.9\;[1.2]  $\cdot$ 10$^{-11}$ & 6.6\;[3.9]  $\cdot$ 10$^{-12}$ & 1.5\;[0.8]  $\cdot$ 10$^{-12}$ \\[0.15cm]
\hline
\end{tabular}
}
\caption{\label{sensMU}
\em Muon collider $2\sigma$ sensitivities on the $\nu_e-\nu_\mu$ TDM. The values in the brackets indicate the sensitivity for a more optimistic scenario, where the background is reduced to a negligible level, and slightly better lepton identification efficiencies are assumed.
}
\end{table}


\begin{figure}
\centering
 \includegraphics[scale=0.45, valign=t]{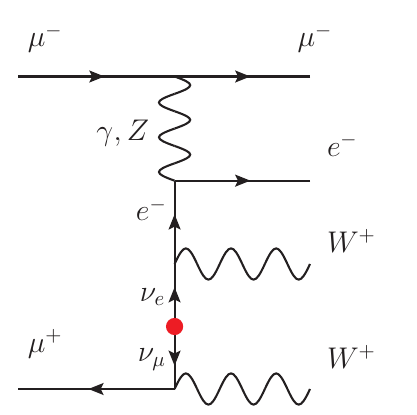}  \includegraphics[scale=0.45, valign=t]{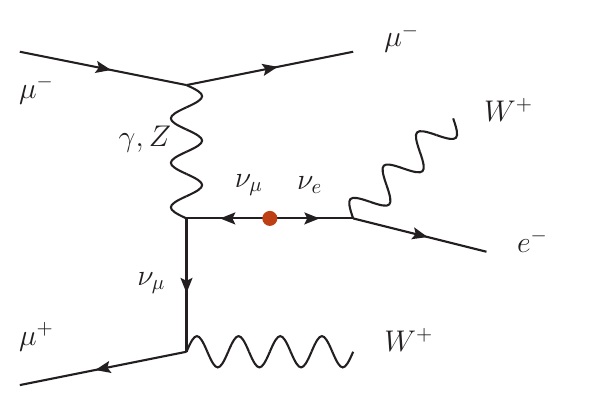} \includegraphics[scale=0.5, valign=t]{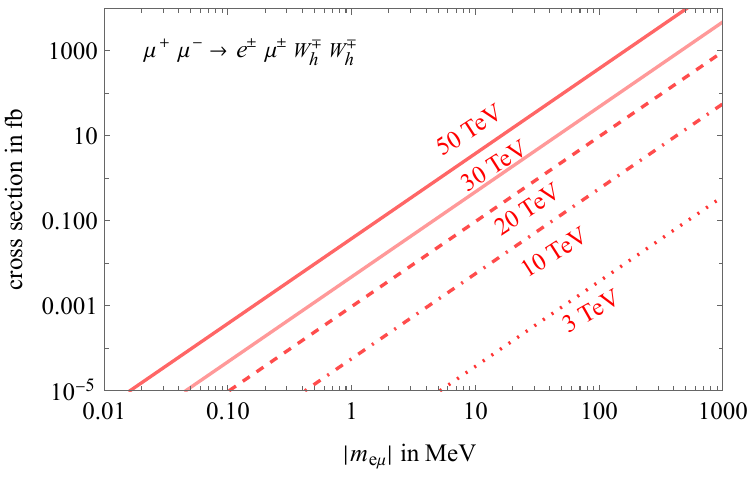}
 \caption{\em Representative leading Feynman diagrams for the $\Delta L=2$ process induced by a Majorana mass $m_{e\mu}$ and the corresponding cross section at a muon collider.}\label{fig:NMM-muCol}
 \end{figure}

\begin{table}[]
\centering
{\small 
\begin{tabular}{|c|ccccc|}
\hline 
 & & & & &  \\
 & & & & &  \\[-0.6cm]
\textsf{$\sqrt{s}$} & 3 TeV & 10 TeV & 20 TeV & 30 TeV & 50 TeV \\
  \hline  & & & & & \\
 & & & & &  \\[-0.6cm]   {$|m_{e\mu}|$}  & 110 [100] MeV & 3.2 [2.5] MeV   &  0.50 [0.31] MeV & 140 [92] keV & 36 [20] keV \\ 
  \hline  & & & & & \\
 & & & & &  \\[-0.6cm]  {$|m_{\mu\mu}|$}  & 300 [140] MeV & 10 [3.5] MeV   &  1.5 [0.44] MeV & 420 [130] keV & 84 [28] keV \\[0.2cm]
\hline
\end{tabular}}\caption{\label{tab:sign-masses}
\em
Muon collider $2\sigma$ sensitivity limits on the Majorana mass matrix entries $|m_{e\mu}|$ and $|m_{\mu\mu}|$. The values in the brackets indicate the sensitivity for a more optimistic scenario (background reduced to a negligible level and slightly better lepton identification efficiencies).} 
\end{table}

\section{Conclusions}

Majorana neutrino TDMs could be efficiently probed at a future muon collider, by analyzing LNV signatures with same-sign dilepton final states. The sensitivity varies from $|\lambda_{e\mu}|\sim 5\cdot 10^{-9}\mu_B$ for energy $\sqrt{s}\simeq 3$ TeV, to $\sim 10^{-12}\mu_B$ for $\sqrt{s}\simeq 50$ TeV, matching the current laboratory bounds for $\sqrt{s}\simeq 30$ TeV. A muon collider would have the unique advantage of a direct and clean identification of lepton number and flavour violation.  
Additionally, the same analysis of LNV signatures at a  muon collider would improve by orders of magnitude (cfr  \cite{Fuks:2020zbm}) the direct bounds on the Majorana neutrino mass matrix entries $m_{e\mu}$ and $m_{\mu\mu}$.

\section{Acknowledgement}
NV thanks the organizers of NOW2024 for the opportunity to present this work. The work of NV is supported by ICSC – Centro Nazionale di Ricerca in High Performance Computing, Big Data and Quantum Computing, funded by European Union – NextGenerationEU, reference code CN\_00000013. MF has received support from the European Union Horizon 2020 research and innovation program under the Marie Sk\l odowska-Curie grant agreements No 860881-HIDDeN and No 101086085–ASYMMETRY.

\end{document}